\documentstyle[11pt]{article}

\newcommand{\be}{\begin{equation}}
\newcommand{\ee}{\end{equation}}
\newcommand{\bea}{\begin{eqnarray}}
\newcommand{\eea}{\end{eqnarray}}

\begin{document}

\begin{titlepage}

\begin{center}

{\Large Centre de Physique Th\'eorique\footnote{Unit\'e Propre de
Recherche 7061} - CNRS - Luminy, Case 907}

{\Large F-13288 Marseille Cedex 9 - France }

\vspace{3 cm}
{\Large {\bf CP VIOLATING METASTABLE STATES AND BARYOGENESIS  IN THE HOT
STANDARD MODEL}}
\vspace{0.3 cm}

{\bf Chris P. KORTHALS ALTES\footnote{email: altes@cpt.univ-mrs.fr}
 and N. Jay WATSON\footnote{email: watson@cptsu4.univ-mrs.fr}}

\vspace{1.5 cm}

{\bf Abstract}

\end{center}

We discuss a novel form of CP violation in the standard model.
It takes place at temperatures
of the order of the electroweak transition, when two
regions with different values of the Wilson line are juxtaposed.
This CP violation is maximal. A sufficient condition is
simply the existence
of a long-lived metastable state; this can occur for fewer than
three generations, and also in the minimal susy standard model.
It leads to baryogenesis in {\it {all}} of these models.
\vspace{2 cm}


\bigskip

\noindent Number of figures: 2

\bigskip

\noindent June 1995

\noindent CPT-94/P.3099

\bigskip

\noindent anonymous ftp or gopher: cpt.univ-mrs.fr

\end{titlepage}

One of the outstanding problems in cosmology is the understanding of
the observed baryon asymmetry of the universe (BAU) \cite{1}.
As the electroweak phase transition is most probably the last occasion
during the cooling of the universe at which this asymmetry may have
come about \cite{2}, it
is only natural to try to use the standard model (SM) to explain
BAU \cite{2,3}. The conventional approach involves the consideration
just below
the critical temperature $T_{ew}$ of the nascent bubbles of normal
vacuum where the Higgs expectation value is non-zero.
The basic idea is that, as the bubbles expand, the scattering of
fermions from the walls differs from that of their CP conjugates as a
result of the SM CP violation encoded in the
Kobayashi-Maskawa \cite{3.5} mixing among
generations. As a consequence, a net baryon-antibaryon separation
builds up. However, actual calculations give conflicting results:
while the authors of Ref.\ \cite{3} concluded that the observed BAU
may just be explained within the SM in this way, the authors
of Ref.\ \cite{4}
concluded that the separation is orders of magnitude too small.
Part of the problem is the smallness of the SM KM CP violation effects.
To solve this problem, many authors have proposed variations
of the SM with a sufficient amount of microscopic CP violation
built in \cite{5}.

However, there exists an independent possible
source of CP violation in the SM
due to thermal effects described by the phase of the Wilson line \cite{6}.
This source of CP violation depends on the existence of very
long-lived metastable minima of the free energy of the universe as a
function of the Wilson line.
These metastable minima are not CP self-conjugate states, and so
break CP invariance spontaneously.
For this source of CP violation to be operational, the only assumption
required is that, rather than falling into the absolute minimum,
the universe has fallen into one of
these long-lived
metastable minima at a temperature somewhere below the GUT
scale (we should however add the caveat that
the thermodynamic admissibility of these
metastable states has been questioned \cite{7,7.5}). The
electroweak transition is supposed to be first order.
In this scenario, the
bubbles of normal vacuum which nucleate at the electroweak phase
transition then have walls that
separate two regions not only of the usual ordered
and disordered Higgs phase
but also of ordered phases parameterized by the Wilson line.

In this letter, we describe both surface and volume effects due to
this alternative scenario. It is shown how the juxtaposed
ordered phases of the Wilson line lead to the localisation
on the wall of fermions, but {\it not} their CP conjugates. This
static, localised
fermion density therefore violates CP invariance maximally.
Also, it is argued that the baryon-antibaryon separation current
arising from the different interactions with the wall
of left- and right-handed fermions
leads to a baryon asymmetry of the required order of
magnitude. Furthermore, these two effects occur not just in the SM,
but in fact are {\em generic} effects occurring in
any model with sufficiently long-lived
CP violating metastable states. Such models include the SM with
{\em fewer} than three
generations and the minimal supersymmetric SM. We wish to emphasize
that this CP violation is spontaneous and
entirely independent of the KM mechanism \cite{3.5}.

The order parameter in hot gauge theories is the Wilson line $\Omega(x)$.
It is defined in terms of the imaginary time formulation of $T \neq 0$
field theory as the path-ordered product of gauge group elements along a
line in the $\tau$ direction:
\be\label{omega}
\Omega(x)
=
{\cal P}\exp{i\int_0^{1\over T}V_0(\vec x,\tau)d\tau}
\ee
where $V_0(\vec x,\tau)$ is the $\tau$ component of the vector potential
of the theory. In the standard model,
the gauge group has SU(3) generators ${\cal A}_0$ with coupling $g_{st}$,
SU(2) generators $A_0$ with coupling $g$ and U(1) generators $B_0$ with
coupling $g^\prime$. In this notation we have \cite{6}
\be\label{v0}
V_0
\equiv
g_{st}{\cal A}_0 + g A_0 + \textstyle{\frac{1}{6}}g^{\prime}B_0.
\ee

{}From the transformation properties of the Wilson line we know that
only its eigenvalues are gauge-invariant. We are thus lead to the
following parametrisation of $V_0$:
\be
{\cal A}_0 = {2\pi T\over {g_{st}}} {\rm diag}
\biggl({q\over 3}+{r\over 2},{q\over 3}-{r\over 2},-{2\over 3}q\biggr),
\hspace{15pt}
A_{0}={2\pi T\over g}{\rm diag}
\biggl({s\over 2},-{s\over 2}\biggr),
\hspace{15pt}
B_0={2\pi T\over g^{\prime}}t.
\ee
The real numbers $q$, $r$, $s$ and $t$ are in the directions of colour
hypercharge, colour isospin, weak isospin and weak hypercharge
respectively. This parametrisation is not gauge-invariant.

$F$, the free energy divided by the factor $\pi^{2}T^{4}$
to be dimensionless,
has been computed as a function of the order parameter
in perturbation theory \cite{6}.
An interesting feature of $F$
is the occurrence of metastable states. These can be present when the
charges of the various particle species are multiples of each other.
This is the case for the weak hypercharges in the standard model.
And indeed, $F$ in the weak hypercharge direction $(0,0,0,t)$
shows two metastable states $t_{1,2} \neq 0$ for two and three
generations (Fig.\ 1), with life times that
exceed by orders of magnitude the advent of the electroweak
transition \cite{6}.
These metastable states have distinct CP conjugates,
as follows from $V_0$ being
a CP-odd quantity.  Thus we have ``spontaneously broken CP invariance''.
The question we wish to address is, how will the
spontaneous CP breaking manifest itself at the electroweak transition?

We consider first surface effects.
The first order character of the transition will show up in
undercooling of the Higgs symmetric phase. Eventually drops of broken
phase
will form. It is easy to show that a non-zero value of the Higgs field
forces the phase of the Wilson line to be either zero or to be in a
metastable minimum of the effective potential built up in the broken
phase \cite{6}. For the purpose of the discussion we will assume that the
former is true. Fig.\ 2 shows the wall of the drop, which consists of the
Higgs profile between $\langle H\rangle\neq 0$ and $\langle H\rangle =0$
and the hypercharge profile between $t = 0$ (i.e.\ $\Omega = 1$) and,
say, $t = t_1$ (i.e.\ $\Omega = {\rm exp}i 2\pi\frac{1}{6} t_{1}$).

Let us look at the Dirac equation around the wall. The wall is given
by the profile, and couples through the $\tau$
component of the covariant derivative. The fermion fields
are antiperiodic. We have
\be\label{dirac}
\Bigl( \gamma_{0}D_0(B) + i\gamma_{z}\partial_{z} \Bigr)\psi(z,\tau)
=
0.
\ee
The covariant derivative is given by
$D_{0}(B)=\partial_{0} + ig'YB_{0}(z)$ for a particle with
hypercharge $Y$.
Using a chiral representation for the $\gamma$'s, $Y$ acts as $Y_R$
($Y_L$) on upper (lower) pairs of components of $\psi$.
We have neglected the Higgs contribution, which is allowed for
fermions with a mass much less than $T_{ew}$.
By the same token, we can look at a fermion with a given handedness,
say the right-handed one.
Our gauge choice in Eq.\ (3) rendered $B_0$
$\tau$-independent, and hence $\psi$ can be Fourier analysed in terms of
$\exp{i(n-{1\over 2})2\pi T}$, $n$ integral.
Then the two normalizable solutions of this equation can be written as
\be
\psi_{1,2}(z,\tau)
=
\biggl( \exp -\gamma_{0}\gamma_{z}\,2\pi T\int^{z}
[ Yt(\zeta) - {\textstyle \frac{1}{2}} ]
d\zeta \biggr)\biggl( \exp -i\pi T\tau \biggr)\phi_{1,2}
\ee
where $n=0$ and $\phi_{1} = (1,0,0,0)^{T}$, $\phi_{2} = (0,0,0,1)^{T}$.
Thus, the hypercharge profile {\it localises} fermions with wavefunctions
proportional to these two spinors. The crucial point is that,
if a given fermion species
localises, its CP conjugate $\gamma_0\gamma_2{\psi}^*$ {\it cannot}.
The residual $\tau$ dependence in the localised wavefunction
disappears in the gauge-invariant combination $\psi^{\dag}\psi$.
Thus, the density is static and localised.

We want to point out that the juxtaposition of two ordered phases
($t = 0$ and $t = t_{1,2}$) is at the root of the localisation process.
The Higgs profile cannot possibly localise fermions, as it separates
ordered and
disordered phases. This is a quite general phenomenon: one needs the
order parameter to be non-zero on both sides of the wall in order to
localise the fermion\footnote{Of course, in the seminal work of
Jackiw and C. Rebbi \cite{9}
the Higgs wall separated ordered phases, and they found
localised modes that were C {\it eigenstates}. As in that case, one may
think here of possible applications in solid state physics.
Physically, there is a comparison between the localised modes and the
``surfactants'' (amphiphyles) of molecular surface physics
(see ref.\ \cite{10}). These surfactants lower the surface tension.}.
Although the numerical contribution of the
localised modes to baryon asymmetry will turn out to be very small,
their presence shows clearly
the CP violating properties of the wall, and how this CP violation is
maximal.

In the context of the strong interactions alone, similar modes have
been found \cite{11}. These modes break C invariance. But their
existence has no physical consequences, since
the strong interaction metastable point \cite{12} destabilises
in the presence of the electroweak forces \cite{6}.

We now turn to volume effects. We will follow the procedure as
laid out in refs.\ \cite{2,3}.
The CP violating wall gives rise to a baryonic
separation current $J_{CP}$. This current sends a net flow of CP
conjugate fermions into the unbroken phase, where baryon
non-conservation \cite{13} is
still effective through the sphaleron activity $f(\rho)$ \cite{3}.
Inside the bubble the sphaleron activity is frozen.
The net baryon number $n_{B}$ is therefore given by
$n_{B} =-J_{CP}f(\rho)$.

The net flow of antiparticles into the unbroken phase
comes from comparing the reflection from the wall
of particles and antiparticles from the outside of the drop, and from
comparing the transmission from the inside to the outside of the drop. We
work in a one dimensional picture\footnote{The phase space
difference between one and three dimensions is not
subject to suppression factors  $(m_s/T)^2$ as in the conventional
approach.} and obtain in the wall restframe \cite{3}
\bea
n_{B}
&=&
{1\over 3}\int_0^{\infty}{d\omega\over {2\pi}}n^v_L(\omega)\Big(
-R^{\dagger}_{LR}R_{LR}
+R^{\dagger}_{\bar R\bar L}R_{\bar R\bar L}
\Big)
f(\rho) + L \leftrightarrow R \nonumber\\
&=&
{1\over 3}\int_0^{\infty}{d\omega\over {2\pi}}
\biggl(n^v_L(\omega)-n^v_R(\omega)\biggr)
R^{\dagger}_{\bar R\bar L}R_{\bar R\bar L}
f(\rho). \label{nb2}
\eea
Here $R_{LR}$ is
the reflection amplitude for right-handed into left-handed
quarks, scattering off the wall from the unbroken phase,
$R_{\bar R\bar L}$ is that of their CP conjugates and
$n^v_L(\omega)$ is the equilibrium distribution for the left-handed
quarks in the wall rest frame,
$n^v_L(\omega)= n[\gamma(\omega+\vec v.\vec p_L)]$.
Eq.\ \ref{nb2} follows from CPT
invariance, which relates amplitudes for particles to
amplitudes for CP conjugates with left and right exchanged.

As is well known \cite{3,4}
the contribution to $n_B$ from transmitted quarks is related
by unitarity and CPT invariance to the same expression
Eq.\ \ref{nb2}, only with $v$ changed into $-v$. For small
enough $v$ one expands in $v$ and retains as lowest order
contribution the quadratic term. One then observes \cite{3,4} that
at one loop order the left- and right-handed
quasiparticles have
different dispersion relations
due to their different interactions with the weak vector bosons.
These dispersion relations are easily obtained
in the presence of the vev in the metastable phase and are
slightly shifted \cite{14}.
It then follows that the integration over $\omega$
in Eq.\ \ref{nb2}  gives a result of
the order of $\alpha_W T$, as in the standard treatment \cite{3,4}.

Now we turn to the difference of matrix elements in Eq.\ \ref{nb2}.
Our  CP violation is spontaneous, and not subject to the GIM
mechanism \cite{14.5}. {}From Fig.\ 1 we see
that the {\it two} generation standard
model has a metastable state, and it easily survives until
the electroweak transition. Decoupling one more generation leaves us
with an even more pronounced dip, and an even longer life time.
So our CP violation does {\it not} require scattering
of one quark generation into another one.

Hence we can draw three important conclusions:
first, from the discussion of the
localised mode we learned that baryon-antibaryon
separation is of order one and hence we gain a factor
$10^{-5}$ in comparison to the traditional CP violation
in the standard model \cite{3};
second, the typical momentum of the quarks in Eq.\ \ref{nb2}
is of order $T\le T_{ew}$
(not of the order of the strange quark mass as with KM
CP violation \cite{3}), resulting in a gain of another factor
of $m_s/T \approx 10^{-3}$;
third, the decoherence mechanism of
Gavela et al.\ \cite{4} is based precisely on multiple scattering
between generations
with a ``coherence length'' of order $(120\,\, {\rm Gev})^{-1}$,
and so does not apply to quarks with momentum of order $T\le T_{ew}$.

Lastly we discuss the sphaleron efficiency factor $f$ in the metastable,
symmetric phase. Its value depends on the typical
number $\rho$ of sphaleron transitions felt by the quark \cite{3}.
This is given by the
combination $\rho= 3D_B \Gamma / v^{2}$
of the diffusion length $D_B$ of the
quarks, the sphaleron rate
$\Gamma=9\Gamma_{sph}/T^3$ and the velocity $v$ of the wall,
all in the symmetric metastable phase\footnote{
This dependence is slightly changed with respect to
the corresponding one in Ref.\ \cite{3}, because our
CP violation extends also to the lepton sector.}.

The metastable phase has SU(2)$\times$U(1) unbroken, like the
stable symmetric phase, since the Higgs vev vanishes.
The non-zero vev of the weak hypercharge potential $B_{0}$
does {\it not} introduce any
other mass scales, even in the zero Matsubara frequency sector.
Had the vev had a component in, say, the weak isospin Lie algebra
$A_{0}$, then the fourth component of the vector potential
would couple as a ``Higgs scalar''
to the spatial components of the gauge fields in that sector and its
vev would have generated a mass.
Obviously this does not happen for the abelian component.
Hence one expects no order of magnitude
difference in the sphaleron rates $\Gamma_{sph}$ in the
metastable and stable symmetric Higgs phases,
i.e.\ $\Gamma_{sph} = (10^{-2} - 1)(\alpha_W T)^4$ \cite{15}.

However the  diffusion lengths and wall velocity may
be much larger in the metastable phase because of the higher free
energy and lower pressure there.
The calculation of the sphaleron activity
was done in the thin wall approximation \cite{3}.
However, the wall has a  typical width \cite{16}
given by the Debye screening length $(g^{\prime} T)^{-1}$,
so necessitates a sufficiently large diffusion length to
guarantee that the quarks penetrate deep enough into the symmetric phase
to feel the sphaleron activity.
For quarks the diffusion length is determined by the
strong interactions and is estimated to be $(4-5)/T$ \cite{3}
in the stable phase, which is of the order of the thickness of the wall.
The uncertainty in the parameter $\rho$
is therefore larger than its counterpart in the stable phase,
leading to a sphaleron activity $f(\rho) \sim 10^{-5} - 1$.

Thus we find a ratio
\be\label{eta}
\eta\equiv {n_B\over s}\sim {10}^{-12} - 10^{-7}
\ee
where $s$ is the 1-d entropy density
taken at the time of the phase transition, $s= {73\over 3}\pi T$.
The first figure for $\eta$ is determined essentially
by the lower limit on the
sphaleron activity, the second figure
by the weak coupling $\alpha_W$ due to the difference in
left- and right-handed dispersion relations,
the sphaleron rate and the number of massless species at $T_{ew}$.
However $s$ increases due to reheating by an order of magnitude
at least. So our ratio diminishes by that same order of magnitude.

Where bubbles coalesce there is a very small matter
contrast due to the zero mode condensate travelling with the wall.
The radius of the bubbles at
coalescence time is of order $R\sim 10^{12}/T$,
so the ratio of the contribution of the
zero modes $\eta_{zm}$ to the ratio Eq.\ \ref{eta}
is of order ${10}^{-12}$, hence a very small contrast.

What if the standard model turns out to be incomplete?
In Fig.\ 1, we
have plotted its minimal SUSY version as well. It is to a very good
approximation just twice as high. The life times of the metastable
states are sufficiently long \cite{14}, so our mechanism works again.

In summary, once part of the universe has fallen into a metastable state
there is baryogenesis of the above order of
magnitude, both for the standard model and its minimal SUSY
extension, consistent with the experimental
number $\eta=(4-6)\times10^{-10}$ \cite{2}.

The metastable states are possible in any theory with a non-trivial
center in the gauge group. If they have a large enough  life time,
CP violation will be of order one at $T_{ew}$, and baryogenesis
will be a {\it{generic}} feature of such theories.

How can part of the universe ``fall''
into the metastable state? The answer is that GUT theories
have metastable states as well (we have checked this for SU(5)
and SUSY-SU(5)). If we assume that these metastable
states are realized at the Planck scale, then it is
energetically possible to fall at $T_{GUT}$ into $t_1$ or $t_2$.
Of course the question is then pushed to the Planck scale,
where, without any microscopic CP violation, the universe
would be divided into equal numbers of causally disconnected
regions of conjugate minima. It may be that microscopic CP
violation plays a role in the choice of metastable state, or
in how the various metastable states develop.

\vspace{10pt}

One of us (C.P.K.A.) wishes to acknowledge the Niels Bohr Institute
and the Danish Research Council for
hospitality and for support, the other (N.J.W.) to acknowledge the
financial support of EC grant ERB4001GT933989.
The comments and criticisms of Rob Pisarski, Kimyeong Lee,
Holger Nielsen, Vladimir Eletsky, Owe Philipsen,
Vadim Kuzmin, Valerie Rubakov and Misha Shaposhnikov have been
very useful.

\pagebreak

{\bf Figure Captions}

\vspace{0.5cm}

Fig.\ 1.\ The normalized free energy $F$ in the weak hypercharge
direction $(0,0,0,t)$. The continuous line is the SM
with three generations, the dashed line the SM with
two generations and the dot-dash line the MSSM.
The Planck free energy $F(0,0,0,0)$ is set to zero.

\vspace{0.5cm}

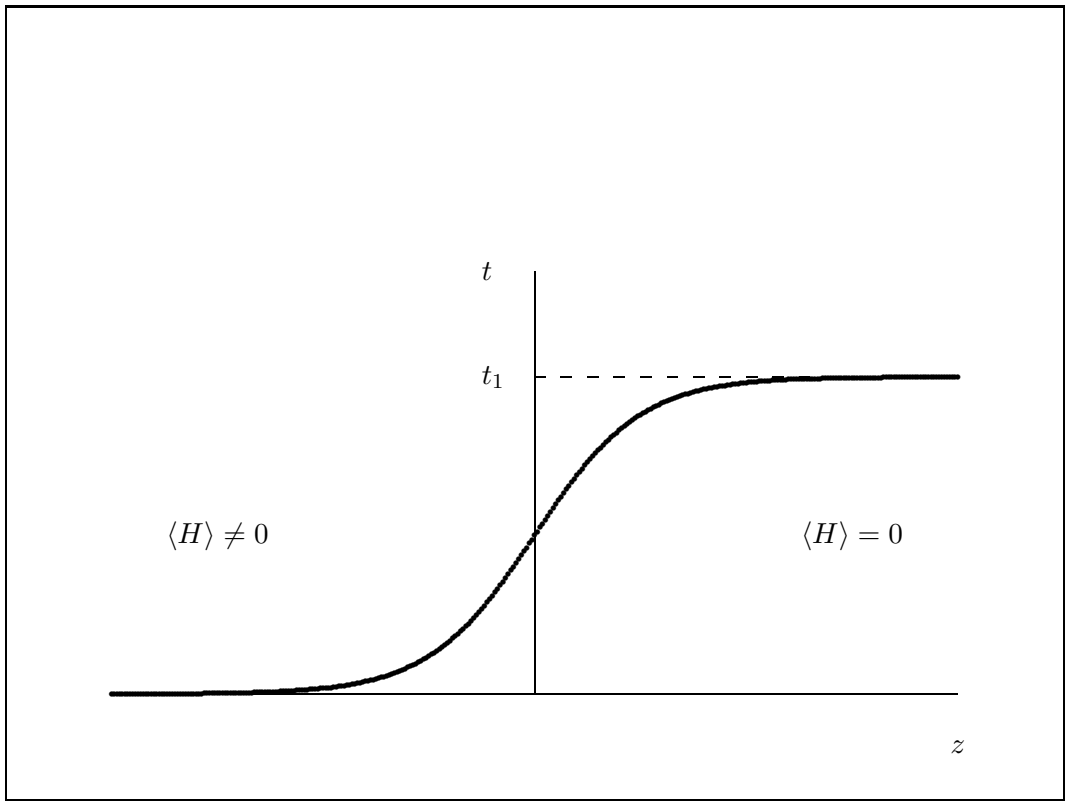
Fig.\ 2.\ The profile of $t \equiv (g'/2\pi T)B_{0}$ and the
Higgs vacuum expectation value in the $z$ spatial
direction, perpendicular to the bubble wall.

\pagebreak

\setlength{\unitlength}{2pt}
\noindent
\begin{picture}(425,220)(0,-100)

\put(20,20){\line(1,0){160}}
\put(100,20){\line(0,1){80}}

\put(180,10){\makebox(0,0)[c]{$z$}}
\put(90,100){\makebox(0,0)[l]{$t$}}
\put(90,80){\makebox(0,0)[l]{$t_{1}$}}
\put(40,50){\makebox(0,0)[c]{$\langle H \rangle\neq 0$}}
\put(160,50){\makebox(0,0)[c]{$\langle H \rangle = 0$}}
\multiput(100,80)(5,0){16}{\line(1,0){2}}

\put(0,0){\framebox(200,150){}}

\put( 20.00, 20.02){\circle*{1.0}}
\put( 20.50, 20.02){\circle*{1.0}}
\put( 21.00, 20.02){\circle*{1.0}}
\put( 21.50, 20.02){\circle*{1.0}}
\put( 22.00, 20.02){\circle*{1.0}}
\put( 22.50, 20.03){\circle*{1.0}}
\put( 23.00, 20.03){\circle*{1.0}}
\put( 23.50, 20.03){\circle*{1.0}}
\put( 24.00, 20.03){\circle*{1.0}}
\put( 24.50, 20.03){\circle*{1.0}}
\put( 25.00, 20.03){\circle*{1.0}}
\put( 25.50, 20.03){\circle*{1.0}}
\put( 26.00, 20.04){\circle*{1.0}}
\put( 26.50, 20.04){\circle*{1.0}}
\put( 27.00, 20.04){\circle*{1.0}}
\put( 27.50, 20.04){\circle*{1.0}}
\put( 28.00, 20.04){\circle*{1.0}}
\put( 28.50, 20.05){\circle*{1.0}}
\put( 29.00, 20.05){\circle*{1.0}}
\put( 29.50, 20.05){\circle*{1.0}}
\put( 30.00, 20.05){\circle*{1.0}}
\put( 30.50, 20.06){\circle*{1.0}}
\put( 31.00, 20.06){\circle*{1.0}}
\put( 31.50, 20.06){\circle*{1.0}}
\put( 32.00, 20.07){\circle*{1.0}}
\put( 32.50, 20.07){\circle*{1.0}}
\put( 33.00, 20.07){\circle*{1.0}}
\put( 33.50, 20.08){\circle*{1.0}}
\put( 34.00, 20.08){\circle*{1.0}}
\put( 34.50, 20.09){\circle*{1.0}}
\put( 35.00, 20.09){\circle*{1.0}}
\put( 35.50, 20.09){\circle*{1.0}}
\put( 36.00, 20.10){\circle*{1.0}}
\put( 36.50, 20.10){\circle*{1.0}}
\put( 37.00, 20.11){\circle*{1.0}}
\put( 37.50, 20.12){\circle*{1.0}}
\put( 38.00, 20.12){\circle*{1.0}}
\put( 38.50, 20.13){\circle*{1.0}}
\put( 39.00, 20.13){\circle*{1.0}}
\put( 39.50, 20.14){\circle*{1.0}}
\put( 40.00, 20.15){\circle*{1.0}}
\put( 40.50, 20.16){\circle*{1.0}}
\put( 41.00, 20.16){\circle*{1.0}}
\put( 41.50, 20.17){\circle*{1.0}}
\put( 42.00, 20.18){\circle*{1.0}}
\put( 42.50, 20.19){\circle*{1.0}}
\put( 43.00, 20.20){\circle*{1.0}}
\put( 43.50, 20.21){\circle*{1.0}}
\put( 44.00, 20.22){\circle*{1.0}}
\put( 44.50, 20.23){\circle*{1.0}}
\put( 45.00, 20.24){\circle*{1.0}}
\put( 45.50, 20.26){\circle*{1.0}}
\put( 46.00, 20.27){\circle*{1.0}}
\put( 46.50, 20.28){\circle*{1.0}}
\put( 47.00, 20.30){\circle*{1.0}}
\put( 47.50, 20.31){\circle*{1.0}}
\put( 48.00, 20.33){\circle*{1.0}}
\put( 48.50, 20.35){\circle*{1.0}}
\put( 49.00, 20.36){\circle*{1.0}}
\put( 49.50, 20.38){\circle*{1.0}}
\put( 50.00, 20.40){\circle*{1.0}}
\put( 50.50, 20.42){\circle*{1.0}}
\put( 51.00, 20.44){\circle*{1.0}}
\put( 51.50, 20.47){\circle*{1.0}}
\put( 52.00, 20.49){\circle*{1.0}}
\put( 52.50, 20.51){\circle*{1.0}}
\put( 53.00, 20.54){\circle*{1.0}}
\put( 53.50, 20.57){\circle*{1.0}}
\put( 54.00, 20.60){\circle*{1.0}}
\put( 54.50, 20.63){\circle*{1.0}}
\put( 55.00, 20.66){\circle*{1.0}}
\put( 55.50, 20.69){\circle*{1.0}}
\put( 56.00, 20.73){\circle*{1.0}}
\put( 56.50, 20.76){\circle*{1.0}}
\put( 57.00, 20.80){\circle*{1.0}}
\put( 57.50, 20.84){\circle*{1.0}}
\put( 58.00, 20.89){\circle*{1.0}}
\put( 58.50, 20.93){\circle*{1.0}}
\put( 59.00, 20.98){\circle*{1.0}}
\put( 59.50, 21.03){\circle*{1.0}}
\put( 60.00, 21.08){\circle*{1.0}}
\put( 60.50, 21.13){\circle*{1.0}}
\put( 61.00, 21.19){\circle*{1.0}}
\put( 61.50, 21.25){\circle*{1.0}}
\put( 62.00, 21.31){\circle*{1.0}}
\put( 62.50, 21.38){\circle*{1.0}}
\put( 63.00, 21.45){\circle*{1.0}}
\put( 63.50, 21.52){\circle*{1.0}}
\put( 64.00, 21.60){\circle*{1.0}}
\put( 64.50, 21.68){\circle*{1.0}}
\put( 65.00, 21.76){\circle*{1.0}}
\put( 65.50, 21.85){\circle*{1.0}}
\put( 66.00, 21.94){\circle*{1.0}}
\put( 66.50, 22.03){\circle*{1.0}}
\put( 67.00, 22.13){\circle*{1.0}}
\put( 67.50, 22.24){\circle*{1.0}}
\put( 68.00, 22.35){\circle*{1.0}}
\put( 68.50, 22.47){\circle*{1.0}}
\put( 69.00, 22.59){\circle*{1.0}}
\put( 69.50, 22.71){\circle*{1.0}}
\put( 70.00, 22.85){\circle*{1.0}}
\put( 70.50, 22.98){\circle*{1.0}}
\put( 71.00, 23.13){\circle*{1.0}}
\put( 71.50, 23.28){\circle*{1.0}}
\put( 72.00, 23.44){\circle*{1.0}}
\put( 72.50, 23.61){\circle*{1.0}}
\put( 73.00, 23.78){\circle*{1.0}}
\put( 73.50, 23.96){\circle*{1.0}}
\put( 74.00, 24.15){\circle*{1.0}}
\put( 74.50, 24.35){\circle*{1.0}}
\put( 75.00, 24.55){\circle*{1.0}}
\put( 75.50, 24.77){\circle*{1.0}}
\put( 76.00, 24.99){\circle*{1.0}}
\put( 76.50, 25.22){\circle*{1.0}}
\put( 77.00, 25.47){\circle*{1.0}}
\put( 77.50, 25.72){\circle*{1.0}}
\put( 78.00, 25.99){\circle*{1.0}}
\put( 78.50, 26.26){\circle*{1.0}}
\put( 79.00, 26.55){\circle*{1.0}}
\put( 79.50, 26.84){\circle*{1.0}}
\put( 80.00, 27.15){\circle*{1.0}}
\put( 80.50, 27.47){\circle*{1.0}}
\put( 81.00, 27.81){\circle*{1.0}}
\put( 81.50, 28.15){\circle*{1.0}}
\put( 82.00, 28.51){\circle*{1.0}}
\put( 82.50, 28.88){\circle*{1.0}}
\put( 83.00, 29.27){\circle*{1.0}}
\put( 83.50, 29.67){\circle*{1.0}}
\put( 84.00, 30.08){\circle*{1.0}}
\put( 84.50, 30.51){\circle*{1.0}}
\put( 85.00, 30.95){\circle*{1.0}}
\put( 85.50, 31.40){\circle*{1.0}}
\put( 86.00, 31.87){\circle*{1.0}}
\put( 86.50, 32.35){\circle*{1.0}}
\put( 87.00, 32.85){\circle*{1.0}}
\put( 87.50, 33.36){\circle*{1.0}}
\put( 88.00, 33.89){\circle*{1.0}}
\put( 88.50, 34.43){\circle*{1.0}}
\put( 89.00, 34.98){\circle*{1.0}}
\put( 89.50, 35.55){\circle*{1.0}}
\put( 90.00, 36.14){\circle*{1.0}}
\put( 90.50, 36.73){\circle*{1.0}}
\put( 91.00, 37.34){\circle*{1.0}}
\put( 91.50, 37.97){\circle*{1.0}}
\put( 92.00, 38.60){\circle*{1.0}}
\put( 92.50, 39.25){\circle*{1.0}}
\put( 93.00, 39.91){\circle*{1.0}}
\put( 93.50, 40.58){\circle*{1.0}}
\put( 94.00, 41.26){\circle*{1.0}}
\put( 94.50, 41.95){\circle*{1.0}}
\put( 95.00, 42.65){\circle*{1.0}}
\put( 95.50, 43.36){\circle*{1.0}}
\put( 96.00, 44.08){\circle*{1.0}}
\put( 96.50, 44.80){\circle*{1.0}}
\put( 97.00, 45.53){\circle*{1.0}}
\put( 97.50, 46.27){\circle*{1.0}}
\put( 98.00, 47.01){\circle*{1.0}}
\put( 98.50, 47.75){\circle*{1.0}}
\put( 99.00, 48.50){\circle*{1.0}}
\put( 99.50, 49.25){\circle*{1.0}}
\put(100.00, 50.00){\circle*{1.0}}
\put(100.50, 50.75){\circle*{1.0}}
\put(101.00, 51.50){\circle*{1.0}}
\put(101.50, 52.25){\circle*{1.0}}
\put(102.00, 52.99){\circle*{1.0}}
\put(102.50, 53.73){\circle*{1.0}}
\put(103.00, 54.47){\circle*{1.0}}
\put(103.50, 55.20){\circle*{1.0}}
\put(104.00, 55.92){\circle*{1.0}}
\put(104.50, 56.64){\circle*{1.0}}
\put(105.00, 57.35){\circle*{1.0}}
\put(105.50, 58.05){\circle*{1.0}}
\put(106.00, 58.74){\circle*{1.0}}
\put(106.50, 59.42){\circle*{1.0}}
\put(107.00, 60.09){\circle*{1.0}}
\put(107.50, 60.75){\circle*{1.0}}
\put(108.00, 61.40){\circle*{1.0}}
\put(108.50, 62.03){\circle*{1.0}}
\put(109.00, 62.66){\circle*{1.0}}
\put(109.50, 63.27){\circle*{1.0}}
\put(110.00, 63.86){\circle*{1.0}}
\put(110.50, 64.45){\circle*{1.0}}
\put(111.00, 65.02){\circle*{1.0}}
\put(111.50, 65.57){\circle*{1.0}}
\put(112.00, 66.11){\circle*{1.0}}
\put(112.50, 66.64){\circle*{1.0}}
\put(113.00, 67.15){\circle*{1.0}}
\put(113.50, 67.65){\circle*{1.0}}
\put(114.00, 68.13){\circle*{1.0}}
\put(114.50, 68.60){\circle*{1.0}}
\put(115.00, 69.05){\circle*{1.0}}
\put(115.50, 69.49){\circle*{1.0}}
\put(116.00, 69.92){\circle*{1.0}}
\put(116.50, 70.33){\circle*{1.0}}
\put(117.00, 70.73){\circle*{1.0}}
\put(117.50, 71.12){\circle*{1.0}}
\put(118.00, 71.49){\circle*{1.0}}
\put(118.50, 71.85){\circle*{1.0}}
\put(119.00, 72.19){\circle*{1.0}}
\put(119.50, 72.53){\circle*{1.0}}
\put(120.00, 72.85){\circle*{1.0}}
\put(120.50, 73.16){\circle*{1.0}}
\put(121.00, 73.45){\circle*{1.0}}
\put(121.50, 73.74){\circle*{1.0}}
\put(122.00, 74.01){\circle*{1.0}}
\put(122.50, 74.28){\circle*{1.0}}
\put(123.00, 74.53){\circle*{1.0}}
\put(123.50, 74.78){\circle*{1.0}}
\put(124.00, 75.01){\circle*{1.0}}
\put(124.50, 75.23){\circle*{1.0}}
\put(125.00, 75.45){\circle*{1.0}}
\put(125.50, 75.65){\circle*{1.0}}
\put(126.00, 75.85){\circle*{1.0}}
\put(126.50, 76.04){\circle*{1.0}}
\put(127.00, 76.22){\circle*{1.0}}
\put(127.50, 76.39){\circle*{1.0}}
\put(128.00, 76.56){\circle*{1.0}}
\put(128.50, 76.72){\circle*{1.0}}
\put(129.00, 76.87){\circle*{1.0}}
\put(129.50, 77.02){\circle*{1.0}}
\put(130.00, 77.15){\circle*{1.0}}
\put(130.50, 77.29){\circle*{1.0}}
\put(131.00, 77.41){\circle*{1.0}}
\put(131.50, 77.53){\circle*{1.0}}
\put(132.00, 77.65){\circle*{1.0}}
\put(132.50, 77.76){\circle*{1.0}}
\put(133.00, 77.87){\circle*{1.0}}
\put(133.50, 77.97){\circle*{1.0}}
\put(134.00, 78.06){\circle*{1.0}}
\put(134.50, 78.15){\circle*{1.0}}
\put(135.00, 78.24){\circle*{1.0}}
\put(135.50, 78.32){\circle*{1.0}}
\put(136.00, 78.40){\circle*{1.0}}
\put(136.50, 78.48){\circle*{1.0}}
\put(137.00, 78.55){\circle*{1.0}}
\put(137.50, 78.62){\circle*{1.0}}
\put(138.00, 78.69){\circle*{1.0}}
\put(138.50, 78.75){\circle*{1.0}}
\put(139.00, 78.81){\circle*{1.0}}
\put(139.50, 78.87){\circle*{1.0}}
\put(140.00, 78.92){\circle*{1.0}}
\put(140.50, 78.97){\circle*{1.0}}
\put(141.00, 79.02){\circle*{1.0}}
\put(141.50, 79.07){\circle*{1.0}}
\put(142.00, 79.11){\circle*{1.0}}
\put(142.50, 79.16){\circle*{1.0}}
\put(143.00, 79.20){\circle*{1.0}}
\put(143.50, 79.24){\circle*{1.0}}
\put(144.00, 79.27){\circle*{1.0}}
\put(144.50, 79.31){\circle*{1.0}}
\put(145.00, 79.34){\circle*{1.0}}
\put(145.50, 79.37){\circle*{1.0}}
\put(146.00, 79.40){\circle*{1.0}}
\put(146.50, 79.43){\circle*{1.0}}
\put(147.00, 79.46){\circle*{1.0}}
\put(147.50, 79.49){\circle*{1.0}}
\put(148.00, 79.51){\circle*{1.0}}
\put(148.50, 79.53){\circle*{1.0}}
\put(149.00, 79.56){\circle*{1.0}}
\put(149.50, 79.58){\circle*{1.0}}
\put(150.00, 79.60){\circle*{1.0}}
\put(150.50, 79.62){\circle*{1.0}}
\put(151.00, 79.64){\circle*{1.0}}
\put(151.50, 79.65){\circle*{1.0}}
\put(152.00, 79.67){\circle*{1.0}}
\put(152.50, 79.69){\circle*{1.0}}
\put(153.00, 79.70){\circle*{1.0}}
\put(153.50, 79.72){\circle*{1.0}}
\put(154.00, 79.73){\circle*{1.0}}
\put(154.50, 79.74){\circle*{1.0}}
\put(155.00, 79.76){\circle*{1.0}}
\put(155.50, 79.77){\circle*{1.0}}
\put(156.00, 79.78){\circle*{1.0}}
\put(156.50, 79.79){\circle*{1.0}}
\put(157.00, 79.80){\circle*{1.0}}
\put(157.50, 79.81){\circle*{1.0}}
\put(158.00, 79.82){\circle*{1.0}}
\put(158.50, 79.83){\circle*{1.0}}
\put(159.00, 79.84){\circle*{1.0}}
\put(159.50, 79.84){\circle*{1.0}}
\put(160.00, 79.85){\circle*{1.0}}
\put(160.50, 79.86){\circle*{1.0}}
\put(161.00, 79.87){\circle*{1.0}}
\put(161.50, 79.87){\circle*{1.0}}
\put(162.00, 79.88){\circle*{1.0}}
\put(162.50, 79.88){\circle*{1.0}}
\put(163.00, 79.89){\circle*{1.0}}
\put(163.50, 79.90){\circle*{1.0}}
\put(164.00, 79.90){\circle*{1.0}}
\put(164.50, 79.91){\circle*{1.0}}
\put(165.00, 79.91){\circle*{1.0}}
\put(165.50, 79.91){\circle*{1.0}}
\put(166.00, 79.92){\circle*{1.0}}
\put(166.50, 79.92){\circle*{1.0}}
\put(167.00, 79.93){\circle*{1.0}}
\put(167.50, 79.93){\circle*{1.0}}
\put(168.00, 79.93){\circle*{1.0}}
\put(168.50, 79.94){\circle*{1.0}}
\put(169.00, 79.94){\circle*{1.0}}
\put(169.50, 79.94){\circle*{1.0}}
\put(170.00, 79.95){\circle*{1.0}}
\put(170.50, 79.95){\circle*{1.0}}
\put(171.00, 79.95){\circle*{1.0}}
\put(171.50, 79.95){\circle*{1.0}}
\put(172.00, 79.96){\circle*{1.0}}
\put(172.50, 79.96){\circle*{1.0}}
\put(173.00, 79.96){\circle*{1.0}}
\put(173.50, 79.96){\circle*{1.0}}
\put(174.00, 79.96){\circle*{1.0}}
\put(174.50, 79.97){\circle*{1.0}}
\put(175.00, 79.97){\circle*{1.0}}
\put(175.50, 79.97){\circle*{1.0}}
\put(176.00, 79.97){\circle*{1.0}}
\put(176.50, 79.97){\circle*{1.0}}
\put(177.00, 79.97){\circle*{1.0}}
\put(177.50, 79.97){\circle*{1.0}}
\put(178.00, 79.98){\circle*{1.0}}
\put(178.50, 79.98){\circle*{1.0}}
\put(179.00, 79.98){\circle*{1.0}}
\put(179.50, 79.98){\circle*{1.0}}
\put(180.00, 79.98){\circle*{1.0}}

\end{picture}

\vspace{25pt}

{\LARGE Figure 2}

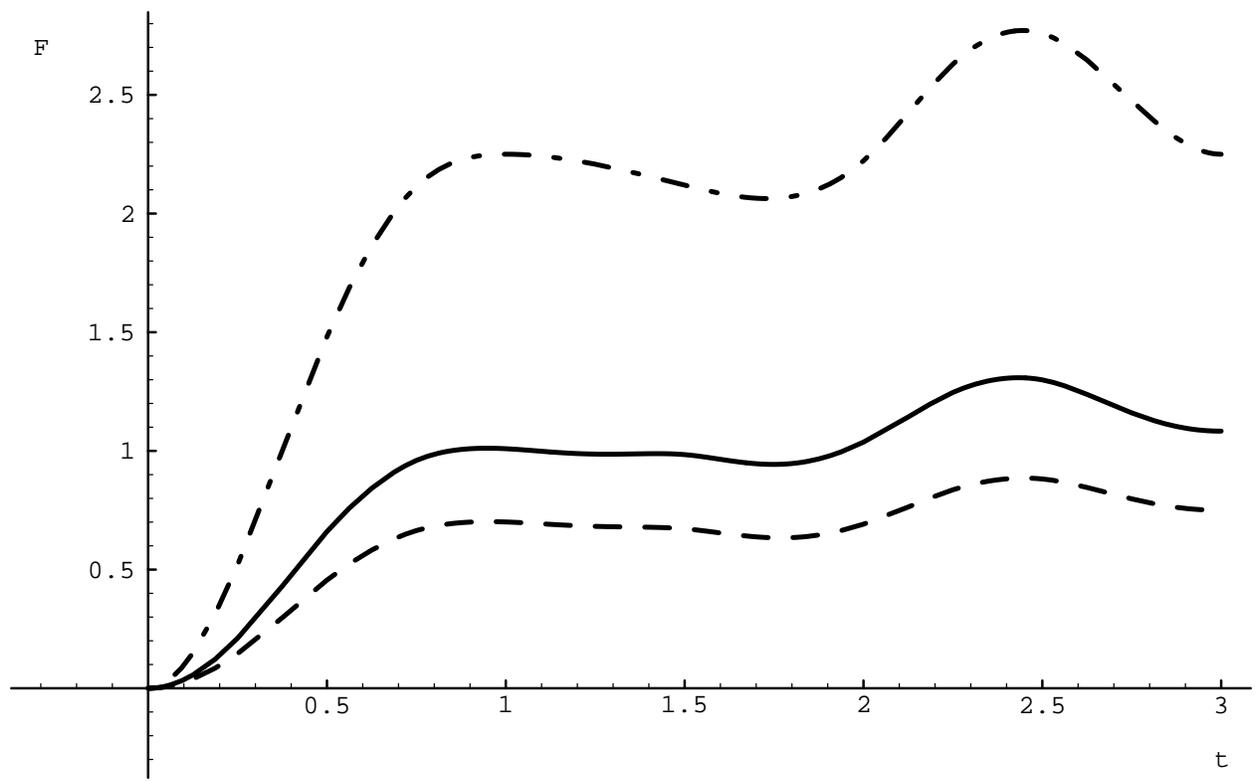
\end{document}